\documentclass[aps,prb,twocolumn,superscriptaddress,showpacs]{revtex4-2}
\usepackage{graphicx}
\usepackage{xcolor}

\usepackage{bm}
\usepackage{braket}

\usepackage{cancel}

\begin{document}

\title{Evolution of valence- and spin-specific local distortions in La$_{2-x}$Sr$_x$CoO$_4$}
\author{J. Okamoto}
\altaffiliation [email: ] {\emph{okamoto.jun@nsrrc.org.tw}} 
\affiliation{National Synchrotron Radiation Research Center, Hsinchu 30076, Taiwan}

\author{A. Chainani}
\affiliation{National Synchrotron Radiation Research Center, Hsinchu 30076, Taiwan}

\author{Z. Y. Chen}
\affiliation{Department of Physics, National Tsing Hua University, Hsinchu 30013, Taiwan}

\author{H. Y. Huang}
\affiliation{National Synchrotron Radiation Research Center, Hsinchu 30076, Taiwan}

\author{A. Singh}
\affiliation{National Synchrotron Radiation Research Center, Hsinchu 30076, Taiwan}

\author{T. Sasagawa}
\affiliation{Materials and Structures Laboratory, Tokyo Institute of Technology, Yokohama, Kanagawa 226-8503, Japan}

\author{D. I. Khomskii}
\affiliation{II. Physikalisches Institut, Universit\"{a}t zu K\"{o}ln, Z\"{u}lpicher Stra{\ss}e 77, D-50937 K\"{o}ln, Germany}

\author{A. Fujimori}
\affiliation{National Synchrotron Radiation Research Center, Hsinchu 30076, Taiwan}
\affiliation{Department of Applied Physics, Waseda University, Tokyo 169-8555, Japan}
\author{C. T. Chen}
\affiliation{National Synchrotron Radiation Research Center, Hsinchu 30076, Taiwan}

\author{D. J. Huang}
\altaffiliation [email: ] {\emph{djhuang@nsrrc.org.tw}} 
\affiliation{National Synchrotron Radiation Research Center, Hsinchu 30076, Taiwan}
\affiliation{Department of Physics, National Tsing Hua University, Hsinchu 30013, Taiwan}

\date{\today}
\begin{abstract}
We present x-ray spectral evidence for the evolution of valence- and spin-specific tetragonal distortions in single-layer cobaltates. Measurements of Co $L_3$-edge resonant inelastic x-ray scattering reveal the $t_{2g}$  electronic structure of Co in hole-doped La$_{2-x}$Sr$_x$CoO$_4$ ($x$ = 0.5, 0.7 and 0.8). As the Sr-doping $x$ increases, the tetragonal splitting of the $t_{2g}$ states of high-spin Co$^{2+}$ decreases, whereas that of low-spin Co$^{3+}$ increases and the population fraction of high-spin Co$^{3+}$ increases. The results enable us to clarify the origin of the change of magnetic anisotropy and in-plane resistivity in the mixed-valence cobaltates caused by the interplay of spin-orbit coupling and tetragonal distortion.
\end{abstract}

%\pacs{75.30.Wx, 71.70.Ch, 78.70.En}

\flushbottom
\maketitle

\thispagestyle{empty}
\section{Introduction}

In strongly correlated $3d$-transition-metal oxides, orderings of the charge, spin, and orbital degrees of freedom compete and cooperate with each other through coupling with the lattice \cite{YTokura2000,MImada1998}. In particular, the orbital degeneracy and its lifting 
due to an external lattice distortion or a Jahn-Teller (JT) effect leads to various electronic or lattice instabilities. For instance, perovskite manganites exhibit various spin orderings and complex phase diagram arising from the orbital degree of freedom \cite{YMnO3, ScMnO3, Mn327}. 
In addition, the spin-orbit coupling in compounds in which the $t_{2g}$ orbitals are partially filled gives rise to a specific change in magnetic anisotropy as the distortions caused by 
JT effect have an effect typically opposite to those due to spin-orbit interaction \cite{Khomskii, Csiszar05, NHollmann2008}. 
Doping holes into a K$_2$NiF$_4$-type antiferromagnetic Mott insulator typically leads to novel phases,
the most striking example being high-transition-temperature cuprate superconductors \cite{JGBednorz86,MImada1998,BKeimer2015}. Single-layer perovskite ``214'' cobaltates La$_{2-x}$Sr$_x$CoO$_4$ are also K$_2$NiF$_4$-type hole-doped Mott insulators and have attracted much attention because they share unusual physical properties with cuprates:  checkerboard-type charge ordering \cite{IAZaliznyak2000,IAZaliznyak2001,ZWLi2016}, hourglass-shaped spin excitations \cite{ATBoothroyd2011,SMGaw2013,YDrees2013,YDrees2014}, and nanoscale-phase separation \cite{YDrees2014,HGuo2015,ZWLi2016p2,ZWLi2016}. Apart from the general features common to many 214 systems, the cobaltates have yet very characteristic properties owing to the existence of several competing spin states of Co$^{3+}$---magnetic high spin (HS) and nonmagnetic low spin (LS) states, which makes the behavior of these systems even richer.

\begin{figure}[t]
\centering
\includegraphics[width=8.5cm]{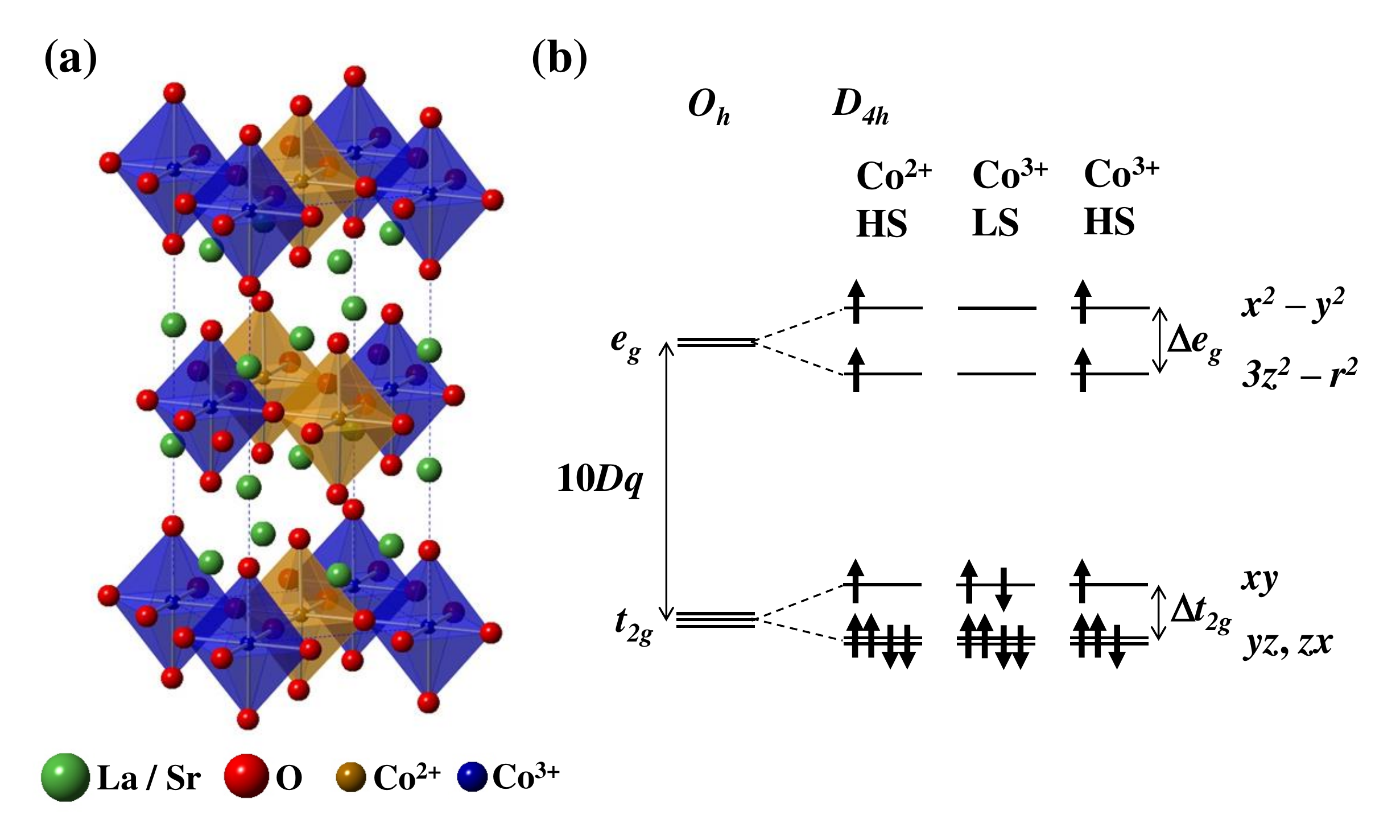}
\caption{\label{D4hCo214} (a) K$_2$NiF$_4$-type lattice structure of La$_{1.5}$Sr$_{0.5}$CoO$_4$ with a checkerboard-type charge ordering. (b) Illustration of the energy levels of Co $3d$ states in a tetragonal $D_{4h}$ crystal field with the CoO$_6$ octahedron elongated along the $c$ axis for HS Co$^{2+}$, LS and HS Co$^{3+}$. These levels are presented in the one-electron picture which excludes the spin-orbit interaction. The tetragonal splittings can be expressed as $\Delta e_g = 4D_{s} +5D_{t}$ and $\Delta t_{2g} = 3D_{s} -5D_{t}$ in terms of parameters $D_s$ and $D_t$ according to the Ballhausen notation \cite{Ballhausen}.}
\end{figure}

The crystal structure of these layered cobaltates is illustrated in Fig.~\ref{D4hCo214}(a).
The competition between crystal-field splitting 10$Dq$ and intra-atomic Hund's exchange interaction $J_{\rm H}$ dictates the spin state of Co ions in La$_{2-x}$Sr$_x$CoO$_4$ \cite{MWHaverkort2006,KTomiyasu2017}. Figure~\ref{D4hCo214}(b) shows the electronic energy levels of Co$^{2+}$ and Co$^{3+}$ under the $D_{4h}$ crystal field according to the one-electron picture.
The $e_{g}$ and $t_{2g}$ states of Co $3d$ are further split by the tetragonal lattice distortion, where the CoO$_6$ octahedra are elongated along the $c$ axis, typical for all layered systems of the 214 type; these splittings are denoted by $\Delta e_g$ and $\Delta t_{2g}$, respectively. For a HS Co$^{2+}$ or Co$^{3+}$ ion, JT distortion yields an additional contribution to the $t_{2g}$ splitting.

La$_{2-x}$Sr$_x$CoO$_4$ is a strong insulator for all Sr doping concentrations; at half doping, $x = 0.5$, it exhibits the highest resistivity \cite{YMoritomo1997} because of checkerboard-type charge ordering of Co$^{2+}$ and Co$^{3+}$,
leading to spin blockade \cite{Maignan04,CFChang2009}. 
In addition, the system exhibits  pronounced magnetic anisotropy with the easy magnetization axis in the $ab$ plane arising from the orbital splitting and spin-orbit coupling for $x$ $<$ 1 \cite{NHollmann2008}, while the easy magnetization axis becomes perpendicular to the $ab$ plane for $x$ = 1.0 \cite{HGuo2016}. 
Various experimental techniques including neutron scattering \cite{IAZaliznyak2000,IAZaliznyak2001,NSakiyama2008,MCwik2009,CTealdi2010,ZWLi2016}, magnetic susceptibility \cite{YMoritomo1997,NHollmann2008,NHollmann2011}, and x-ray absorption spectroscopy (XAS) \cite{CFChang2009} have been applied to study the electronic structures that underly the phase diagram of 
La$_{2-x}$Sr$_x$CoO$_4$. The elucidation of the change 
of the spin states and electronic properties of Co$^{2+}$ and Co$^{3+}$ ions in response to the change of hole doping remains, however, a challenging task, in order to reveal the origin of the doping-dependent transport and magnetic anisotropy. To clarify this question is the main goal of the present investigation.

Resonant inelastic x-ray scattering (RIXS) is an effective spectroscopic method to measure excitations of correlated-electron materials with charge, spin, orbital and lattice degrees of freedom \cite{LJPAment2011}. A series of Co $L_3$-edge RIXS measurements recently enabled us to unravel the valence and spin states of Co oxides in relation to their physical properties \cite{MMvanSchooneveld2012,KTomiyasu2017,HNiwa2017,YYokoyama2018,RPWang2018,RPWang2019}. 
Here we present Co $L_{3}$-edge RIXS of La$_{2-x}$Sr$_x$CoO$_4$ to investigate the Co spin states and the tetragonal distortion by comparing the measured RIXS spectra with cluster-model calculations. The results indicate that the tetragonal distortion of the Co$^{3+}$O$_6$ octahedra and the HS population of Co$^{3+}$ increase with Sr doping, whereas those of Co$^{2+}$O$_6$ decrease. Concomitant change of the spin state of Co$^{3+}$ was also observed.

\section{Experimental}

Single crystals of La$_{2-x}$Sr$_x$CoO$_4$ were grown by the floating-zone method \cite{YOkimoto2013}. The crystals were cut and polished for $x$~=~0.5 and naturally cleaved for $x$~=~0.7 and 0.8  with the sample surface in the $ab$ plane. We recorded Co $L_{2,3}$-edge XAS spectra at 25 K using the partial-fluorescence-yield method at beamline 08B of Taiwan Light Source (TLS) of National Synchrotron Radiation Research Center (NSRRC) in Taiwan. The XAS energy resolution was $\sim$~0.3 eV.  XAS spectra of La$_{2-x}$Sr$_{x}$CoO$_4$ were recorded with two geometries: normal incidence ($I_{\perp}$) in which the electric-field vector $\boldsymbol{\varepsilon}$ of the incident beam perpendicular to the $c$ axis and off normal incidence ($I^{\prime}$) in which the angle between 
$\boldsymbol{\varepsilon}$ and the $a$ axis is 60$^{\circ}$. 
That is, XAS spectra with $\boldsymbol{\varepsilon} \parallel c$ and isotropic XAS spectra are $I_{\parallel} = \frac{4}{3}(I^{\prime} - \frac{1}{4}I_{\perp})$, and $\frac{1}{3}(2I_{\parallel} +I_{\perp})$, respectively. 
We performed Co $L_3$-edge RIXS with an AGM-AGS spectrometer \cite{CHLai2014} 
at TLS beamline 05A. The total energy resolution of RIXS was 90 meV; the base pressure of the RIXS chamber was 1$\times10^{-8}$ torr. Samples were cooled to 20 K with liquid helium. The measurement geometry is illustrated in Fig. \ref{XASandRIXS}(b). The scattering plane was in the $ac$ plane of the sample crystal, and the incident and scattering angles were set to 20$^{\circ}$ and 90$^{\circ}$, respectively. The electric-field vector $\boldsymbol{\varepsilon}$ of the incident beam was parallel and perpendicular to the scattering plane for $\pi$  and  $\sigma$ polarizations, respectively. Under this scattering geometry with $\pi$ polarization, the elastic intensity is suppressed and the projection of the electric-field vector $\boldsymbol{\varepsilon}$ onto the $c$ axis is enhanced.

\section{Results and Analysis}
\subsection{Co $L_{2,3}$-edge XAS}

\begin{figure}[t]
\centering
\includegraphics[width=8.5cm]{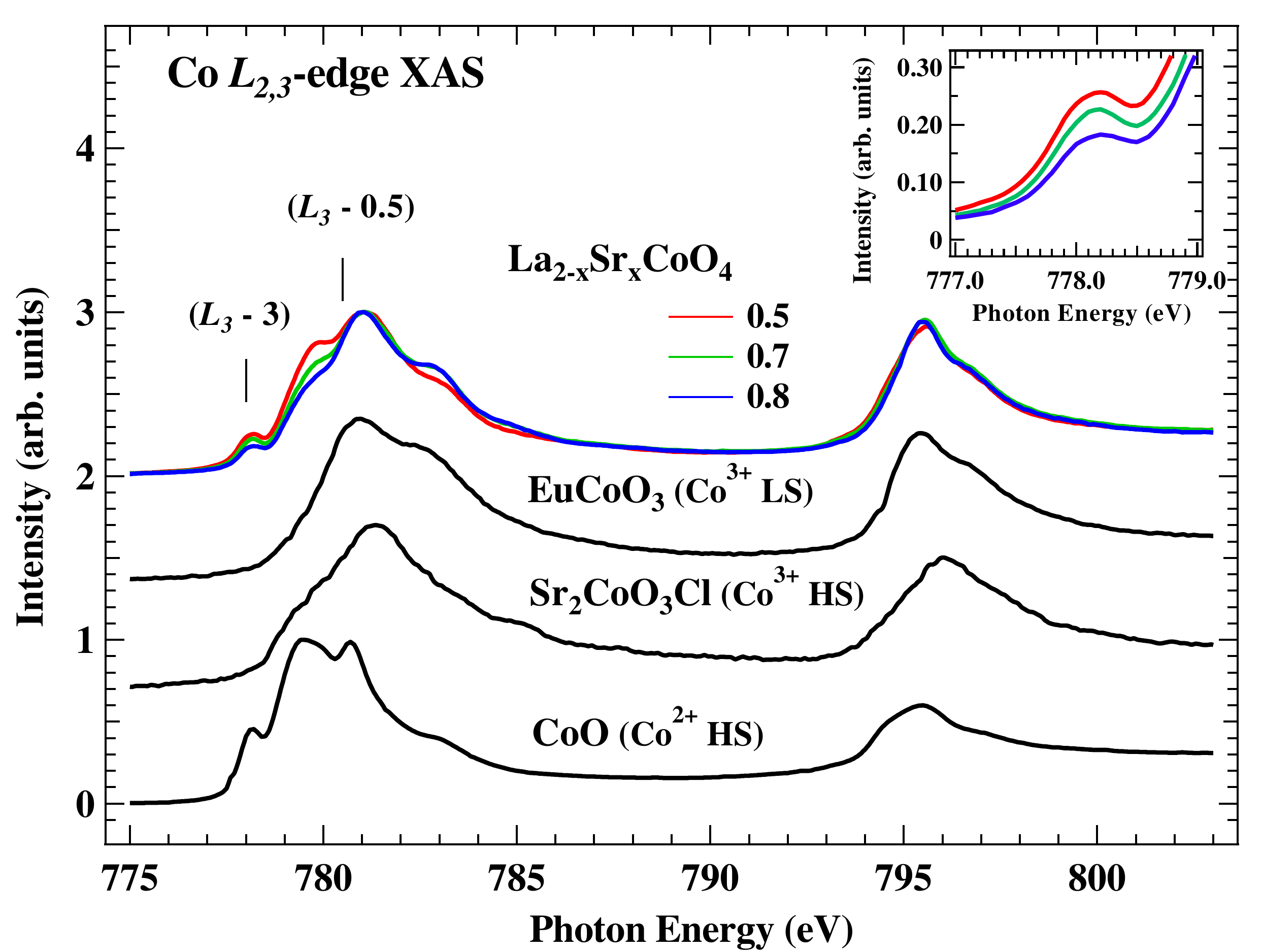}
\caption{\label{CoXAS} Isotropic Co $L_{2,3}$-edge XAS spectra of La$_{2-x}$Sr$_{x}$CoO$_4$ ($x$ = 0.5, 0.7, and 0.8) recorded at 25 K. XAS spectra of CoO, Sr$_2$CoO$_3$Cl, and EuCoO$_3$ are plotted as the reference spectra of HS Co$^{2+}$, HS Co$^{3+}$, and LS Co$^{3+}$, respectively. The vertical bars indicate photon energies 778 eV $(L_{3}-3)$ and 780.5 eV $(L_{3}-0.5)$, respectively. The inset shows a zoom-in view of the XAS spectra of La$_{2-x}$Sr$_{x}$CoO$_4$ around $(L_3-3)$.}
\end{figure}

Figure \ref{CoXAS} shows the isotropic Co $L_{2,3}$-edge XAS spectra of La$_{2-x}$Sr$_{x}$CoO$_4$ ($x$ = 0.5, 0.7, and 0.8) and three reference compounds: Sr$_2$CoO$_3$Cl, EuCoO$_3$ and CoO, which contain HS Co$^{3+}$, LS Co$^{3+}$, and HS Co$^{2+}$ ions, respectively. Co $L_{2,3}$-edge XAS spectra of La$_{1.5}$Sr$_{0.5}$CoO$_4$ with $\boldsymbol{\varepsilon}\perp c$ and $\boldsymbol{\varepsilon} \parallel c$ are consistent with those reported in Ref. \cite{CFChang2009}. There exist four pronounced features in the XAS spectra of La$_{2-x}$Sr$_{x}$CoO$_4$.  We focus on two features marked with vertical bars at 778 eV denoted by $(L_{3}-3)$ and 780.5 eV denoted by $(L_{3}-0.5)$ to unravel the contributions of HS Co$^{2+}$ and LS Co$^{3+}$. The XAS intensity at $(L_{3}-3)$ changes proportionally to the Co$^{2+}$ population ($1-x$) as shown in the inset of Fig. \ref{CoXAS}. Among the reference samples, CoO has a much stronger XAS feature at $(L_{3}-3)$ than others; this XAS feature is derived mostly from HS Co$^{2+}$.  In addition, all the reference XAS spectra show a strong intensity at $(L_{3}-0.5)$, indicating that a separation of the contributions of HS Co$^{2+}$, LS Co$^{3+}$, and HS Co$^{3+}$ states to the XAS feature of La$_{2-x}$Sr$_{x}$CoO$_4$ at $(L_{3}-0.5)$ is not straightforward. 

\subsection{Co $L_{3}$-edge RIXS}

\begin{figure*}[t]
\centering
\includegraphics[width=17cm]{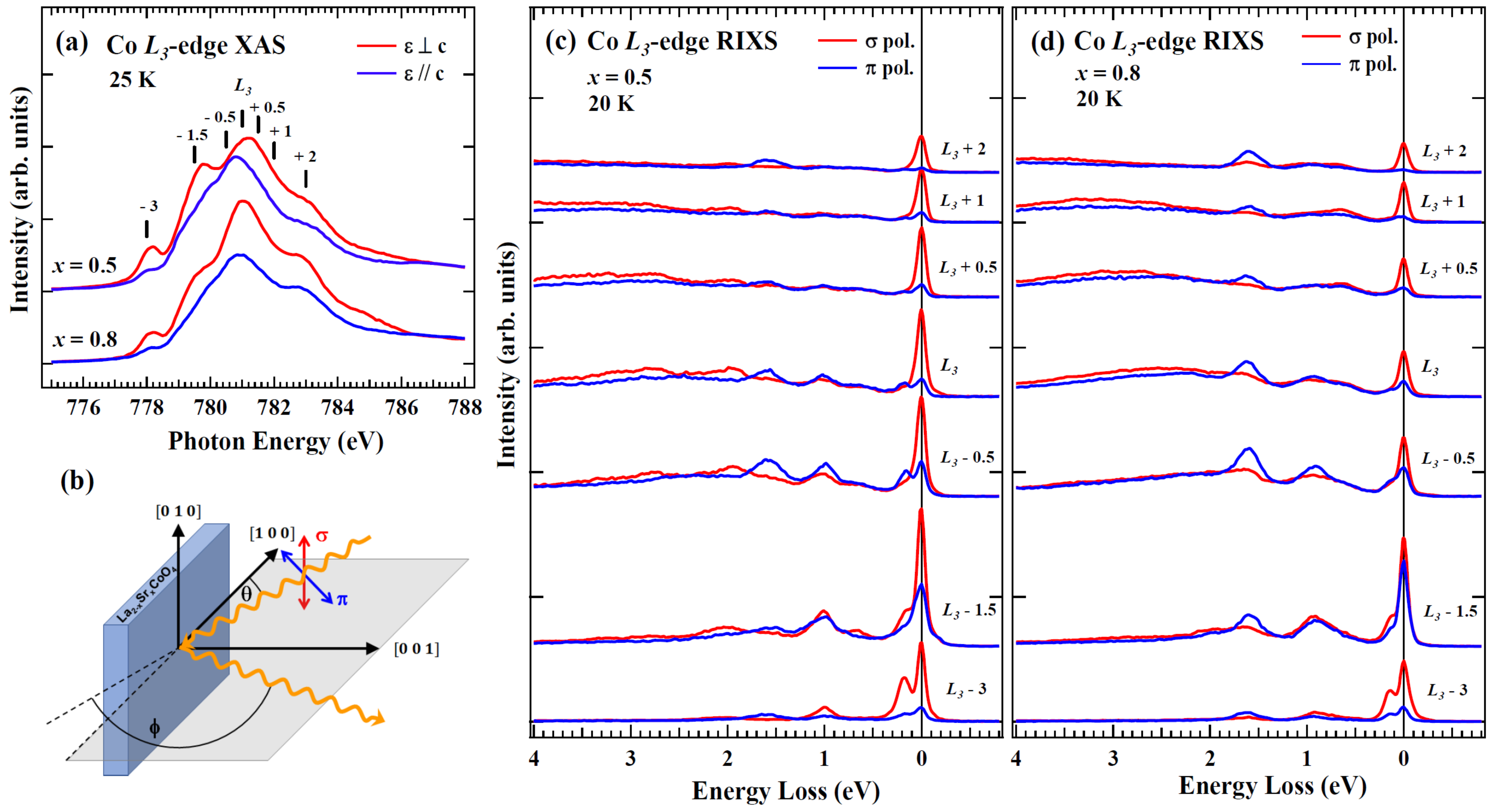}
\caption{\label{XASandRIXS}Co $L_{3}$-edge XAS and RIXS spectra of La$_{2-x}$Sr$_{x}$CoO$_4$ ($x$ = 0.5 and 0.8). (a) XAS spectra recorded at 25 K with $\boldsymbol{\varepsilon}\perp c$ and $\boldsymbol{\varepsilon}\parallel c$. (b) Illustration of the RIXS experiment geometry for $\sigma$ and $\pi$ polarizations. Co $L_3$-edge RIXS spectra of (c) $x$ = 0.5 and (d) $x$ = 0.8 measured at 20 K excited with $\sigma$ and $\pi$ polarized x-rays at various energies from 778 eV ($L_3-3$) to 783 eV ($L_3+2$) shown by vertical bars in (a). RIXS spectra are normalized to the incident beam intensity.}
\end{figure*}

To elucidate the electronic structures of Co$^{2+}$ and Co$^{3+}$ in La$_{1.5}$Sr$_{0.5}$CoO$_4$, we resorted to Co $L_3$-edge RIXS measurements of La$_{2-x}$Sr$_x$CoO$_4$ ($x$ = 0.5 and 0.8) as shown in Figs. \ref{XASandRIXS}(c) and \ref{XASandRIXS}(d) with the incident photon energies indicated in Fig. \ref{XASandRIXS}(a). Because fluorescence overlaps with RIXS structures of energy loss above 2 eV when the incident x-ray energy is above ($L_3-0.5$), we focus on RIXS structures of energy loss below 2 eV. There exist two distinct RIXS features at energy loss of 0.2 eV and 1.6 eV; their intensities vary with incident photon energy and Sr doping. 
The 0.2-eV RIXS structure is pronounced at the incident photon energy ($L_3 - 3$) for $\sigma$ polarization; it is also discernible at the incident photon energy ($L_3 - 3$) for $\pi$ polarization. With the XAS spectra shown in Fig. \ref{XASandRIXS}(a) and calculated XAS of HS Co$^{2+}$ \cite{CFChang2009} taken into account, we found that the observed 0.2-eV RIXS is derived from HS Co$^{2+}$. 
In addition, the 1.6-eV RIXS structure is enhanced for $x$ = 0.8, particularly when the incident photon energy is above ($L_3 - 0.5$). This observation implies that the 1.6-eV RIXS structure is derived from Co$^{3+}$. We performed RIXS calculations to analyze the measured RIXS spectra and to investigate the local electronic structures of Co$^{2+}$ and Co$^{3+}$ in La$_{2-x}$Sr$_x$CoO$_4$.

\subsubsection{RIXS of high-spin Co$^{2+}$}

\begin{figure}[t]
\centering
\includegraphics[width=8.5cm]{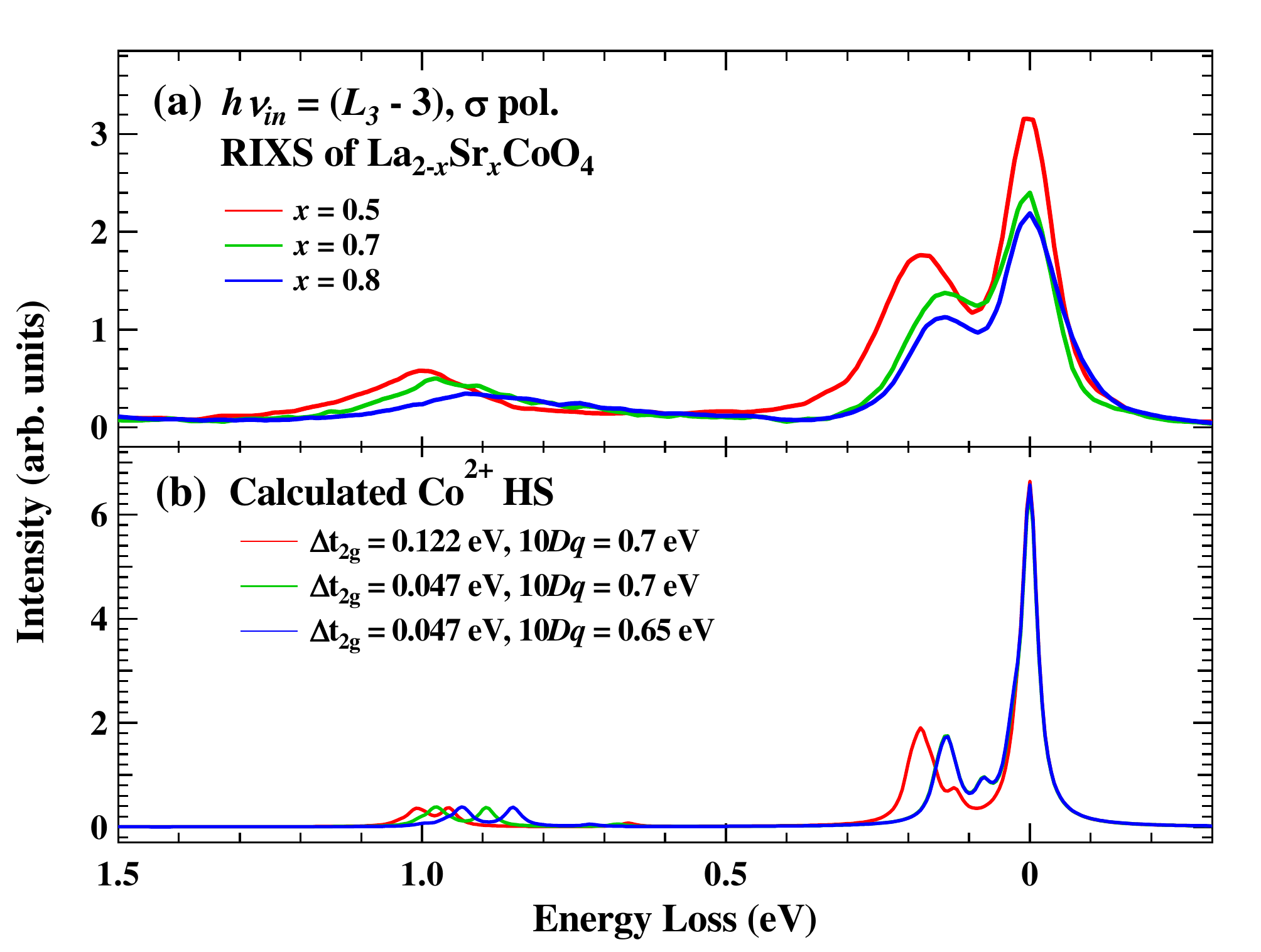}
\caption{\label{L3m3} Co $L_3$-edge RIXS spectra of La$_{2-x}$Sr$_x$CoO$_4$ excited with $\sigma$-polarized x rays of energy 778 eV denoted by $(L_{3}-3)$. 
(a) RIXS spectra recorded at 20 K normalized to the incident beam intensity. (b) Calculated RIXS spectra for HS Co$^{2+}$ of tetragonal symmetry with various parameter values for the tetragonal distortion $\Delta t_{2g}$ and crystal field 10$Dq$.}
\end{figure}   

Figure \ref{L3m3}(a) depicts Co RIXS spectra excited with $\sigma$-polarized x rays of energy set to $(L_{3}-3)$  for samples with $x$ = 0.5, 0.7, and 0.8. In addition to the elastic scattering, there exist distinct RIXS features at energies about 1~eV and 0.2~eV. The former is derived from electronic transitions of Co$^{2+}$ from the HS ground state $t_{2g}^{5}e_{g}^{2}$  of symmetry $^{4}T_{1g}$ to a HS excited state $t_{2g}^{4}e_{g}^{3}$ of symmetry $^{4}T_{2g}$; its excitation energy is the energy required to promote a $t_{2g}$ electron to an $e_{g}$ orbital, i.e., 10$Dq$. 
The 0.2-eV excitation arises from the transitions of Co$^{2+}$ within the $t_{2g}$ manifold; its excitation energy is thus affected by the tetragonal distortion but is insensitive to 10$Dq$. Measurements of these RIXS excitations lead us to determine unambiguously the values of $10Dq$ and $\Delta t_{2g}$.  We simulated the RIXS spectra of HS Co$^{2+}$ with a configuration-interaction CoO$_6$ cluster model that includes the full atomic multiplets and the hybridization with the O $2p$ ligands \cite{Quanty,Paracalc}. The calculated spectra plotted in Fig. \ref{L3m3}(b) also show two distinct excitations at energies near 0.2 eV and 1 eV, in good agreement with the RIXS measurements. 
The measured spectrum is best described using parameters $10Dq$~=~0.7~eV and $\Delta t_{2g}$~= 122~meV for HS Co$^{2+}$ at half doping. In addition, the change of Sr doping $x$ results in a shift of the excitations at 0.2 eV and 1 eV to lower energies indicating the change of the Co$^{2+}$ electronic structure. Comparison of these spectra with calculations shows that both the tetragonal splitting $\Delta t_{2g}$ and the crystal field 10$Dq$ decrease with $x$ increasing from 0.5 to 0.8, i.e., $\Delta t_{2g}$ being changed from 122~meV to 47~meV and 10$Dq$ from 0.7~eV to 0.65~eV. These results reveal that, for HS Co$^{2+}$, the tetragonal distortion of the CoO$_6$ octahedra decreases, reflecting that the in-plane Co-O bond length increases with Sr doping $x$. 

\subsubsection{RIXS of low-spin and high-spin Co$^{3+}$}

\begin{figure}
\centering
\includegraphics[width=8cm]{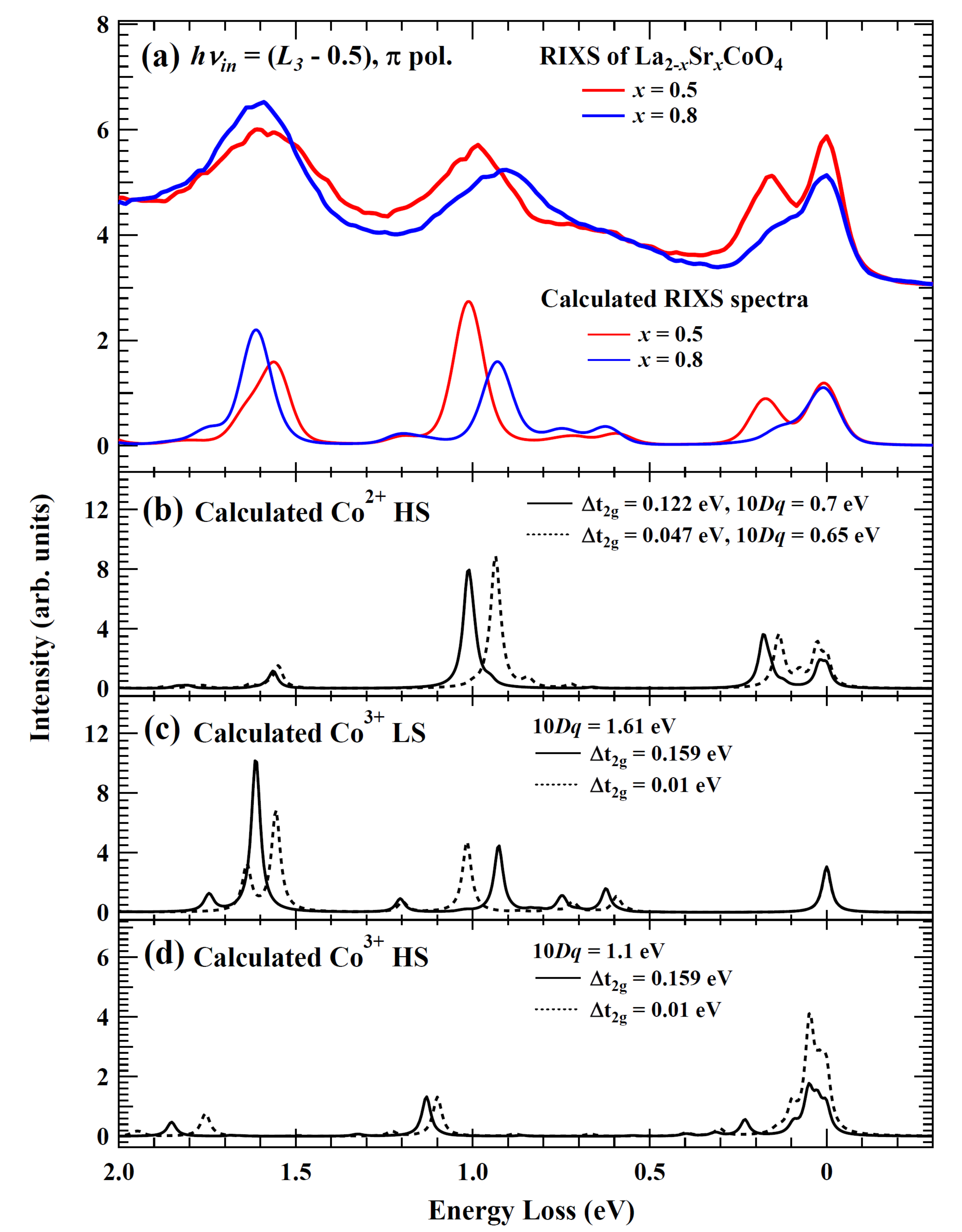}
\caption{\label{L3m05} (a) Measured Co $L_3$-edge RIXS spectra of La$_{2-x}$Sr$_x$CoO$_4$ ($x$ = 0.5 and 0.8) with incident photon energy $(L_3-0.5)$ and $\pi$ polarization. Calculated RIXS spectra for $x$ = 0.5 and 0.8 are obtained from the sum of the calculated RIXS spectra of (b) HS Co$^{2+}$, (c) LS Co$^{3+}$, and (d) HS Co$^{3+}$  with estimated populations plotted in Fig. \ref{DsandP} (a). A spectral broadening has been applied to the summed spectra.}
\end{figure}

In La$_{2-x}$Sr$_{x}$CoO$_4$, Co$^{3+}$ can be in either the LS or the HS state. For large $10Dq$, the electronic configuration energetically favors the LS state, whereas the HS state is stabilized with increased $J_{\rm H}$ or decreased $10Dq$.
Co $L_3$-edge RIXS measurements have been demonstrated to be an effective method
to probe the spin state of Co$^{3+}$ in LaCoO$_3$ \cite{KTomiyasu2017}.  
Here, we further examine the local electronic structure of Co$^{3+}$ in La$_{2-x}$Sr$_{x}$CoO$_4$ and its spin state by tuning the incident photon energy to ($L_3-0.5$). 
Figure \ref{L3m05}(a) shows the RIXS spectra of La$_{2-x}$Sr$_{x}$CoO$_4$ with incident x rays of $\pi$ polarization to suppress the elastic scattering. The structures above 2 eV overlap with fluorescence as shown in Figs. \ref{XASandRIXS}(c) and \ref{XASandRIXS}(d). Figure \ref{L3m05}(b) plots calculated RIXS spectra of HS Co$^{2+}$ with the same parameters as those used in Fig. \ref{L3m3}(b). They reproduce excitations of 0.2 eV and 1 eV. 
As Sr doping is increased, the 0.2-eV feature is strongly suppressed, consistent with the above assignment that the structure at 0.2 eV mainly comes from HS Co$^{2+}$. 
Similar to the RIXS measurements of LaCoO$_{3}$ that we reported previously \cite{KTomiyasu2017}, clear but broad excitation features at 1 eV and 1.6 eV are spectral evidence for the LS ground state of Co$^{3+}$ and are attributed to the $t_{2g}$-to-$e_g$ transitions in the one-electron picture. 
The 1.6-eV excitation is derived from a transition from the low-spin $^{1}A_{1g}$ ground state to another low-spin state of symmetry $^{1}T_{1g}$ without spin flip. 
Its excitation energy depends on the combined effect of $10Dq$, $p$-$d$ hybridization, $p$-to-$d$ charge-transfer energy, the effective onsite Coulomb energy $U_{dd}$ \cite{KTomiyasu2017}, and the tetragonal distortion parameter $D_t$, but is less sensitive to the other tetragonal distortion parameter $D_s$. 
The 1-eV excitation  involves spin flip, resulting from transitions to the intermediate-spin (IS) $^{3}T_{2g}$ states. Its excitation energies decrease with increased tetragonal distortion. 
The 0.6-eV excitation involves spin flip, resulting from transition to the other IS $^3T_{1g}$ states.

Figures \ref{L3m05}(c) and \ref{L3m05}(d) show calculated RIXS spectra of LS and HS Co$^{3+}$ ions, respectively.  We used parameter values typical for Co$^{3+}$ compounds \cite{Paracalc}. The $10Dq$ and $\Delta t_{2g}$ of LS Co$^{3+}$  are 1.61~eV and 10~meV, respectively, for $x=0.5$.  The calculated RIXS spectra of LS Co$^{3+}$ reproduce the features at 1 eV and 1.6 eV of the measured spectra, corroborating the aforementioned symmetry explanation of excited states. In contrast, the calculated RIXS spectrum of HS Co$^{3+}$ is dominated by elastic scattering and excitations of energy less than 0.1 eV and fails to account for the measurements. This observation indicates that the Co$^{3+}$ ions in the checkerboard-type charge ordering of La$_{1.5}$Sr$_{0.5}$CoO$_4$ are in the LS state, in agreement with the results of neutron scattering \cite{IAZaliznyak2000} and XAS \cite{CFChang2009}, which explain the extremely insulating nature of La$_{1.5}$Sr$_{0.5}$CoO$_4$ according to the spin-blockade mechanism. For $x$ = 0.8, a larger $\Delta t_{2g}$($\sim$ 160 meV) is needed to explain the shifts of the RIXS peaks at 1 and 1.6 eV, as shown in Figs. \ref{L3m05}(a) and \ref{L3m05}(c).

\section{Discussion}

The combined results of RIXS measurements and cluster-model calculations reveal that the LS Co$^{3+}$ ion has a larger crystal-field splitting $10Dq$ than the HS Co$^{2+}$ ion; this explains the low-spin character of Co$^{3+}$, which leads to a much smaller ionic size and shorter Co-O bonds. In addition, both CoO$_6$ octahedra of HS Co$^{2+}$ and LS Co$^{3+}$ show a tetragonal distortion in which the apical Co-O bond is elongated along the $c$ axis owing to the tetragonal lattice structure \cite{MCwik}. The $t_{2g}$ splitting $\Delta t_{2g}$ defined in Fig. \ref{D4hCo214}(b) for both cases is positive. Furthermore, the JT effect on the HS Co$^{2+}$ modifies the $t_{2g}$ splitting, whereas LS Co$^{3+}$ is JT inactive. We found that the $\Delta t_{2g}$ value of HS Co$^{2+}$ at half doping is larger than that of LS Co$^{3+}$: 122~meV and 10~meV, respectively. 

For heavily hole-doped La$_{2-x}$Sr$_{x}$CoO$_4$, the effective magnetic moment per Co ion is found to be substantially decreased from  3.87 $\mu_{\rm B}$  at $x$ = 0.5 to $\approx$ 2.6 $\mu_{\rm B}$ at $x$ = 0.8, 
accompanied by a steep decrease of the in-plane resistivity \cite{YMoritomo1997} and suppression of magnetic anisotropy \cite{NHollmann2008}. In this regime, the checkerboard charge order breaks down; there exist incommensurate charge correlations \cite{ZWLi2016}.  These substantial changes of magnetic and electronic properties indicate the partial conversion of LS Co$^{3+}$ into HS Co$^{3+}$, which might lead to incommensurate spin correlations, and a large $p$-$d$ hybridization.  

\begin{table}
\caption{\label{L3m05Int} Ratio of integrated intensities of RIXS structures of La$_{1.2}$Sr$_{0.8}$CoO$_4$ against La$_{1.5}$Sr$_{0.5}$CoO$_4$ with the incident photons of energy $(L_3-0.5)$  and $\pi$ polarization.}
\begin{ruledtabular}
\begin{tabular}{lllll}
 & Elastic & 0.2-eV & 1-eV & 1.6-eV \\
\hline
$I_{x=0.8}/I_{x=0.5}$ & 0.83$\pm$0.05 & 0.52$\pm$0.05 & 0.87$\pm$0.05 & 1.2$\pm$0.05
\end{tabular}
\end{ruledtabular}
\end{table}

\begin{figure}
\centering
\includegraphics[width=8cm]{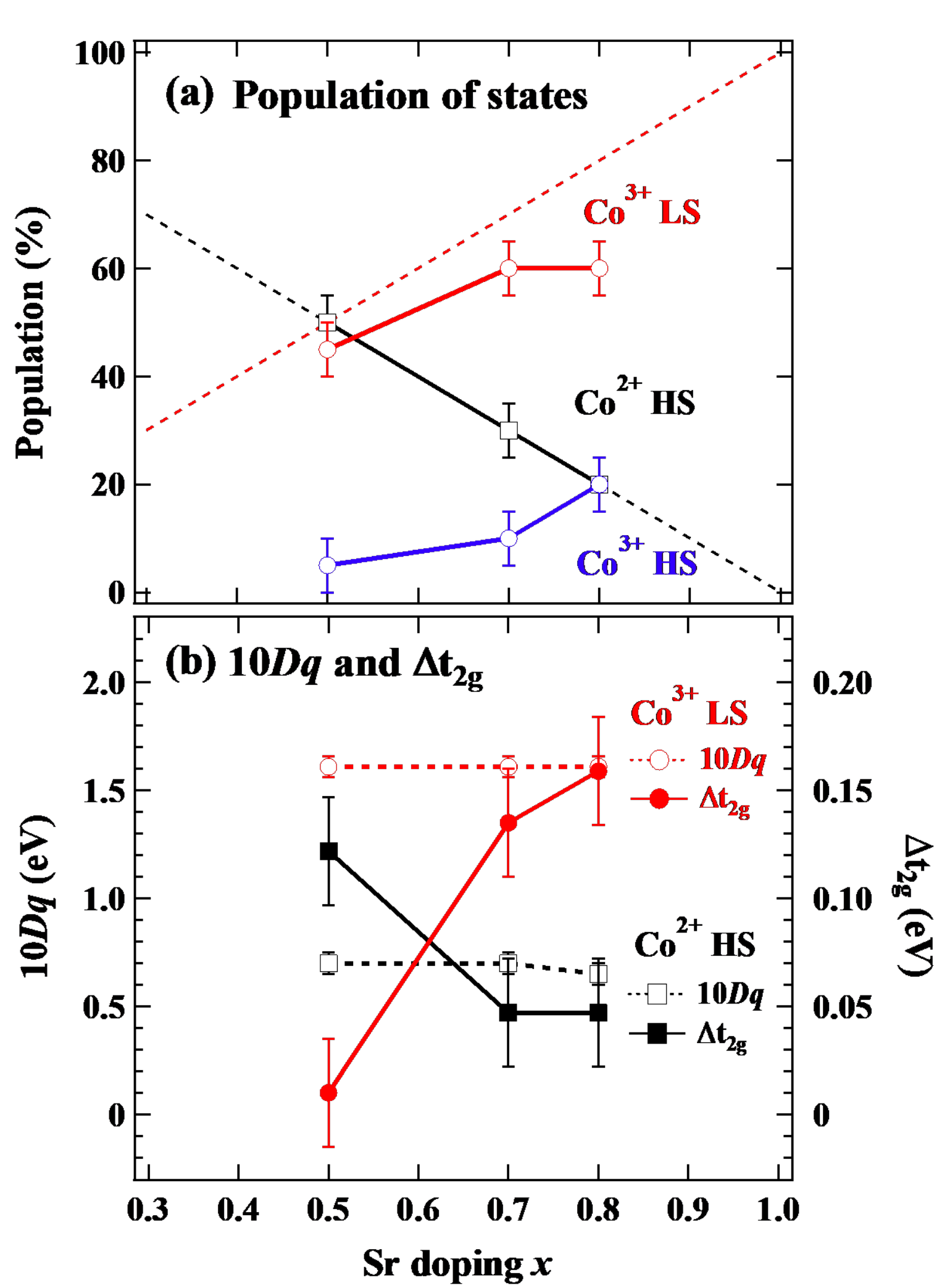}
\caption{\label{DsandP} Sr-doping dependence of (a) population of states estimated with RIXS results and (b) crystal field and tetragonal distortion of CoO$_6$. Black (red) broken lines in (a) are nominally expected Co$^{2+}$ (Co$^{3+}$) populations, respectively.}
\end{figure}

We also used RIXS to investigate the doping dependence of the Co spin state. The intensity changes of the RIXS features shown in Fig.~\ref{L3m05}(a) provide a measure of the population changes of LS Co$^{3+}$ and HS Co$^{2+}$, as plotted in Fig. \ref{DsandP}(a). Black (red) broken lines are nominally expected Co$^{2+}$ (Co$^{3+}$) populations, respectively. The 0.2-eV RIXS excitation arises from transitions within the $t_{2g}$ manifold of HS Co$^{2+}$; its intensity change reflects the change in the population of HS Co$^{2+}$. When the Sr doping $x$ is increased from 0.5 to 0.8, the intensities of the 0.2-eV and the 1.6-eV RIXS features decrease by 48\% and increase by 20\%, respectively, revealing that the population of LS Co$^{3+}$ increases to 60\% and that of HS Co$^{2+}$ decreases to 20\%. This suggests that the population of HS Co$^{3+}$ increases from $\sim$0 to $20$\%. Furthermore, the intensity of quasielastic scattering is decreased by only 17\%, smaller than the observed 48\% decrease of the 0.2-eV RIXS intensity, providing evidence for the existence of HS Co$^{3+}$ in heavily doped La$_{2-x}$Sr$_{x}$CoO$_4$. These intensity changes are summarized in Table \ref{L3m05Int}. From the population changes of LS Co$^{3+}$ and HS Co$^{2+}$, the populations of HS Co$^{3+}$ for $x=$~0.7 and 0.8 are estimated to be 10\% and 20\%, respectively. Calculated RIXS spectra with the estimated populations of HS Co$^{2+}$, LS Co$^{3+}$, and HS Co$^{3+}$ reproduce the measured RIXS spectra as shown in Fig. \ref{L3m05}(a). Our RIXS results indicate a significant population of the HS Co$^{3+}$ state in heavily Sr-doped  La$_{2-x}$Sr$_x$CoO$_4$, in agreement with the evidence from XAS results \cite{ZWLi2016}.

Figure \ref{DsandP}(b) shows that hole-doping dependences of the $t_{2g}$ splittings of HS Co$^{2+}$ and LS Co$^{3+}$ show opposite trends. $\Delta t_{2g}$ of LS Co$^{3+}$ increases from 10~meV to 159~meV for $x$ altered from 0.5 to 0.8. This significant increase in $\Delta t_{2g}$ reflects the substantially decreased in-plane lattice constants in the highly doped regime. As the Sr concentration is increased, there is a tendency for the doped holes of Co $3d$ to move predominantly in the $ab$ plane to gain kinetic energy, and consequently contraction in the $ab$ plane is favorable. 
The contraction in the $ab$ plane also increases the in-plane $pd\pi$ hybridization between the $t_{2g}$ and ligand $p$ orbitals, contributing to a decrease in the in-plane resistivity \cite{YMoritomo1997}.

We can also use the change of $\Delta t_{2g}$ to estimate the variation of the Co-O bond length in response to the Sr doping. 
For $x$ = 0.5, the in-plane and apical Co-O bond lengths in the unit of {\AA} are estimated, respectively, to be 1.954 and 2.190 for Co$^{2+}$, and 1.888 and 2.075 for Co$^{3+}$ by neutron scattering \cite{MCwik}. 
In other words, the ratio $R$ of the Co-O bond lengths Co-O$_{\rm apical}$/Co-O$_{\rm in-plane}$ is 1.121 for Co$^{2+}$ and 1.099 for Co$^{3+}$ at half doping $x$ = 0.5. 
Assuming that the change of $R$ of Co$^{2+}$ and Co$^{3+}$ in the same crystal structure is proportional to the change of $\Delta t_{2g}$, one can estimate the $R$ for different Sr dopings. We find that, for $x$ = 0.5, a reduction of 112 meV in $\Delta t_{2g}$ corresponds to a decrease of 0.022 in $R$. That is, $R$ decreases from 1.121 ($x$ = 0.5) to 1.106 ($x$ = 0.7 and 0.8) for the Co$^{2+}$ site, and increases from 1.099 ($x$ = 0.5) to 1.124 ($x$ = 0.7) and 1.128 ($x$ = 0.8) for the Co$^{3+}$ site. If we further assume that the in-plane Co-O bond lengths for the Co$^{2+}$ and Co$^{3+}$ are nearly the same for various hole dopings, we estimate the apical Co-O bond length of the Co$^{2+}$ and Co$^{3+}$ sites as shown in Table \ref{CoObond}. Note that the apical bond lengths of the Co$^{3+}$ sites are overestimated; the average Co-O bond length of the Co$^{3+}$ site at $x$ = 1 is 2.061 \AA (apical) and 1.902 \AA (in-plane) \cite{MCwik}. This overestimation is expected to come from the increase of HS Co$^{3+}$ population. HS Co$^{3+}$ becomes stable not only by the increase of tetragonal distortion but also by the decrease of crystal field 10$Dq$, which leads to the elongation of the “in-plane” Co-O bond length and the suppression of $R$.

\begin{table}
\caption{\label{CoObond} Estimated Co-O bond-length ratio $R$ = Co-O$_{\rm apical}$/Co-O$_{\rm in-plane}$ and the apical Co-O bond lengths for the Co$^{2+}$ and Co$^{3+}$ sites with Sr doping from the change of $\Delta t_{2g}$. $R$ and Co-O$_{\rm apical}$ at $x$ = 0.5 are taken from Ref. \cite{MCwik}. The Co-O$_{\rm in-plane}$ distances at $x$ = 0.5 are 1.954 {\rm \AA}  and 1.888 {\rm \AA} for Co$^{2+}$ and Co$^{3+}$, respectively \cite{MCwik}.}
\begin{ruledtabular}
\begin{tabular}{lllll}
  &  & $x$ = 0.5 & $x$ = 0.7  & $x$ = 0.8 \\
\hline
Co$^{2+}$ & $\Delta t_{2g}$ (eV) & 0.122 & 0.047 & 0.047  \\
               & $R$              & 1.121 & 1.106  & 1.106  \\
               & Co-O$_{\rm apical}$ (\AA) & 2.190 & 2.162 & 2.162  \\
\hline
Co$^{3+}$ & $\Delta t_{2g}$ (eV) & 0.010 & 0.135  & 0.159  \\
               & $R$              & 1.099 & 1.124 & 1.128   \\
			 & Co-O$_{\rm apical}$ (\AA) & 2.075 & 2.121 & 2.130  
\end{tabular}
\end{ruledtabular}
\end{table}
 
The observed evolution of the tetragonal distortion explains the change of magnetic anisotropy from the in-plane easy axis at half doping \cite{NHollmann2008} to the out-of-plane easy axis in the limit of $x=1$ \cite{HGuo2016}. Because the spin-orbit coupling of Co $3d$ is of the same order of magnitude as the $t_{2g}$ splitting, a single-electron state in the partially filled $t_{2g}$ shell can have an unquenched orbital moment. In this scheme, the $t_{2g}$ states are better expressed in terms of complex linear combinations of $\ket{xy}$, $\ket{yz}$, and $\ket{zx}$. Wave functions denoted $\ket{d^x_{\pm1}}$, $\ket{d^y_{\pm1}}$, and $\ket{d^z_{\pm1}}$ with an effective orbital momentum $\widetilde{l}=1$ can be formed for the orbital moment quantized along axes $x$, $y$, and $z$, respectively \cite{NHollmann2008,Khomskii,effective_orb}. 
At half doping, the magnetic anisotropy is dictated by HS Co$^{2+}$, which has one hole in the $t_{2g}$ orbitals. Because of the inherent elongation of the CoO$_6$ octahedron in the tetragonal lattice structure and the spin-orbit interaction, a $t_{2g}$ hole originally in the $\ket{xy}$ orbital with no magnetic anisotropy acquires the $\ket{d^x_{\pm1}}$ or $\ket{d^y_{\pm1}}$ orbital character and an orbital moment in the $ab$ plane becomes energetically favorable \cite{Csiszar05,NHollmann2008,Khomskii}. HS Co$^{2+}$ consequently exhibits a magnetic anisotropy in that the in-plane susceptibility $\chi_{ab}$ is larger than the out-of-plane one $\chi_c$. One can similarly argue that the HS Co$^{3+}$ favors an elongated CoO$_6$ octahedron and a spin along the $c$ axis. There is one hole in the degenerate $\ket{yz}$ and $\ket{zx}$ orbitals, as illustrated in Fig.~\ref{D4hCo214}(b). If the spin-orbit interaction is included, this hole has an unquenched orbital moment in the $z$ direction through the formation of  $\ket{d^z_{\pm1}}$, leading to an easy magnetization axis along the $c$ axis.
To verify the above arguments,  we compared the energy change of HS Co$^{3+}$ under an external magnetic field by using atomic multiplet calculations. The energy of HS Co$^{3+}$ under a field of 1 T along the $z$ axis is lower than that without an external field by 0.21 meV;  it, however, remains unchanged if the field is along the $x$ axis. Therefore, a magnetic anisotropy of $\chi_{c} > \chi_{ab}$ is expected in the heavily hole-doped regime, consistent with the magnetic-susceptibility measurements of LaSrCoO$_4$ \cite{HGuo2016}.

\section{Summary}

By tuning the resonant energy, we exploited RIXS to probe the electronic structure of transition-metal ions of specific valence in a mixed-valence compound. The crystal-field and tetragonal-distortion energies of Co$^{2+}$ and Co$^{3+}$ are separately obtained. The Co$^{3+}$ ions gradually change from the low-spin to a high-spin state with increasing Sr concentration.  The crystal-field splittings $10Dq$ of LS Co$^{3+}$  and HS Co$^{2+}$ are 1.61~eV and 0.7~eV at half doping, respectively.
The tetragonal $t_{2g}$ splittings of HS Co$^{2+}$ and LS Co$^{3+}$ have opposite trends in the hole-doping dependence. For the doping $x$ altered from 0.5 to 0.8, $\Delta t_{2g}$ of LS Co$^{3+}$ increases from 10~meV to 159~meV, while that of  HS Co$^{2+}$ decreases from 122~meV to 47~meV.
These results unravel the underlying mechanism of the change of magnetic anisotropy and in-plane resistivity of La$_{2-x}$Sr$_x$CoO$_4$ through the interplay between the tetragonal distortion and the spin-orbit coupling.

\section*{Acknowledgements}

We thank the NSRRC staff for technical help during our RIXS measurements at the AGM-AGS beamline of Taiwan Light Source. We thank C. F. Chang and A. C. Komarek for valuable discussions. This work was supported in part by the Ministry of Science and Technology of Taiwan through MOST 109-2112-M-213-010-MY3 and also by JSPS through KEKENHI No. 19K03741. D.I.K. was supported by the Deutsche Forschungsgemeinschaft (DFG, German Research Foundation), project number 277146847 – CRC 1238.

\end{document}